# Microstructure and charge-ordering transitions in LuFe$_2$O$_4$

Y. Zhang, H. X. Yang , C. Ma, H. F. Tian & J. Q. Li *

Beijing National Laboratory for Condensed Matter Physics, Institute of Physics,

Chinese Academy of Sciences, Beijing 100080, P. R. China.

Microstructure properties, phase transitions, and charge ordering in the LuFe$_2$O$_4$ materials have been extensively investigated by means of transmission electron microscopy (TEM). The (001) twins as a common defect frequently appear in the LuFe$_2$O$_4$ crystals along the c-axis direction, and the crystals across each boundary are rotated by 180° with respect to one another. The in-situ cooling TEM observations reveal remarkable temperature dependence of the superstructures in correlation with charger ordering (CO). The Fe$^{2+}$, Fe$^{3+}$, and Fe$^{+2.5}$ ions are found to be crystallized in ordered stripes at the frustrated ground state characterized by a modulation with the wave vector of $\mathbf{q_1}$ =(1/3 1/3 2). In-situ heating TEM observations on LuFe$_2$O$_4$ clearly demonstrate that this modulation becomes evidently invisible above a critical temperature of about Tc=530K. These facts suggest that the CO should be the essential driving force for the structural transitions and ferroelectricity observed in present system.

PACS number(s): 77.84.-s; 71.27.+a; 61.14.-x; 71.23.An

# Ⅰ. INTRODUCTION

In recent years, extensive investigations of the charge-frustrated $RFe_2O_4$ system (R=Y, Er, Yb, Tm and Lu) revealed a rich variety of interesting physical properties, such as ferroelectricity, magnetodielectric, spin and charge ordering, and certain works have also focused on their immense potential for technical applications [1-8]. Actually, the ferroelectricity in $RFe_2O_4$ system has been studied for decades, it has a large dielectric constant with notable Debye-type frequency dependence [3,4]. Recently, the giant room-temperature magnetodielecric response was observed in the $LuFe_2O_4$ single crystalline material [5], this distinctive property is considered as a significant feature for developing a new generation of multifunctional devices for microelectronics [6-8]. It is also noted that the $LuFe_2O_4$ material undergoes sequential phase transitions with lowering temperature as revealed by means of x-ray or neutron diffractions [9-11]. Moreover, the anomalous dielectric behaviors become often visible in correlation with these phase transitions [12], it is believed that the presence of CO and the charge/spin frustration have evident effects on the remarkable physical properties observed in these materials [13,14]. In the structural point of view, $RFe_2O_4$ material belongs to the rhombohedral system (R-3m) and consists of two typical layers, the hexagonal double layers of Fe ions with an average valence of $Fe^{2.5+}$ are sandwiched by thick $R_2O_3$ layers [15-17]. Though previous literatures made certain attempts to clarify the correlation between the $Fe^{2+}/Fe^{3+}$ order and the ferroelectricity in $RFe_2O_4$ materials [18], the detailed structural evolution in connection with the CO modulation as well as ferroelectric phase transition are not well understood at this time. In this paper, we will report on the microstructure features and common defects in $LuFe_2O_4$ materials as revealed by means of transmission electron microscopy ( TEM ), then we have also carried out in-situ TEM observation in a wide temperature range of 20K-550K, structural phase transitions, especially the CO process, have been extensively analyzed in details.

## Ⅱ. EXPERIMENTAL

The polycrystalline $LuFe_2O_4$ materials used in present study were prepared by conventional solid-state reaction under controlled oxygen partial pressure using $CO_2$-$H_2$ mixture at 1200℃ for 48h. Structural and phase purity characterizations were carried out by the powder x-ray diffraction (XRD) using a RIGAKU diffractometer with Cu Kα radiation. Microstructure and chemical composition were analyzed on a Philips XL30 scanning electron microscope. Transmission electron microscopy (TEM) investigations were performed on a Tecnai F20 field-emission electron microscope operating at a voltage of 200kV. This TEM machine is equipped with both low-temperature (down to 20K) and high temperature (up to 1200K) holders. TEM samples were prepared by mechanical polishing, dimpling and subsequently ion milling.

## Ⅲ. RESULTS AND DISCUSSION

Fig. 1a shows the XRD pattern taken from a $LuFe_2O_4$ sample at room temperature. All diffraction peaks can be well indexed by the hexagonal cell with the space group of R-3m and lattice parameters of a=3.444Å and c=25.259 Å, no peaks from impurity phase are observed. Fig. 1b shows a schematic structural model of the $LuFe_2O_4$ crystal, clearly illustrating the alternative sequence of Lu-O layer and the Fe-O double layers stacking along the c axis direction. This layered structural feature could primarily affect the growth of $LuFe_2O_4$ crystal. Fig. 1c shows a SEM image for a typical sample showing the notable layered structural features. Our careful structural examinations on the samples grown under slightly different conditions suggest that the $LuFe_2O_4$ crystal in general grows following the layer-by-layer mode.

Figs. 2a-d show the selected-area electron diffraction patterns taken along several relevant zone-axis directions at room temperature, respectively. The reflection condition for the main diffraction spots is in good agreement with the space group of R-3m, it generally is written as $-H+K+L=3n$ for (HKL). In addition to the main reflections, superstructure reflections often appears at the systematic positions of (1/3 1/3 L) as prolonged streaks along the c axis direction [9,19]. Our extensive TEM observations demonstrated that this notable superstructure depends evidently on temperature in a large temperature range as discussed in following context.

Fig. 3 shows a HRTEM image taken along the [100] zone axis direction, clearly displaying the atomic structural feature in the $LuFe_2O_4$ crystal. This image was obtained from the thin region of a crystal under the defocus value at around -45nm. The metal atom positions are recognizable as white dots, and the atomic layers of the $LuFe_2O_4$ phase along the c direction can be clearly read out. Image calculations, based on the proposed structure model of fig. 1b was carried out by varying the crystal thickness from 2 to 5 nm and the defocus value from –60 to 60 nm. A calculated image with the defocus value of -45nm and the thickness of 3nm, superimposed on the image, appears to be in good agreement with the experimental one.

The common defect in $LuFe_2O_4$ materials as observed in our TEM investigations is the structural twinning along the c-axis direction, Fig. 4a shows a bright-field TEM image illustrating the twining domains in a $LuFe_2O_4$ crystal. Contrast anomalies at the twin boundaries directly suggest the presence of crystal defects and local strains in association with these planar defects. These twinning planes in general are parallel to the basic *a-b* plane. Fig. 4b-d show the corresponding electron diffraction patterns taken from three representative areas indicated in fig. 4a, illustrating the structural evolution across a boundary. Diffraction patterns from area I and area III have a well-defined twinning relationship as clearly demonstrated in the fig.4c. Hence, these structural lamellae can be well interpreted as 180° twins, and the $LuFe_2O_4$ crystals cross each boundary are rotated by 180° with respect to one another. Fig. 5a is a [100] zone-axis HRTEM image at higher magnification, clearly showing the atomic structural feature at a twin boundary. Careful examinations on the atomic structure in vicinal regions of a boundary suggest that stacking faults often appear at this kind of boundaries. For instance, Fig. 5b shows a HRTEM image for a twin boundary, careful analysis and theoretical simulation for this boundary reveal a layered sequence of LuO-FeO-FeO-FeO-FeO-LuO in contrast with the atomic sequence of LuO-FeO-FeO-LuO-FeO-FeO-LuO in $LuFe_2O_4$ crystal along c axis direction.

The most significant structural features concerned in present system is about the superstructure modulation in connection with the CO which plays the key role for understanding the remarkable ferroelectricity observed in this kind of materials.

Recently, we have carried out a series of experimental investigations on the structural properties of LuFe$_2$O$_4$ and focused on the charge ordering and low-temperature phase transitions as partially reported in ref. [20]. The in-situ cooling TEM observations demonstrated that the CO in the ground state of LuFe$_2$O$_4$ could be well characterized by two structural modulations at low temperatures.

Fig. 6a shows the [1-10] zone diffraction pattern taken at about 20K, demonstrating the presence of clear superstructure spots following the main diffraction spots. These superstructure spots can be assigned respectively to two structural modulations (**q$_1$** and **q$_2$**) as clearly illustrated in the schematic pattern of Fig. 6b. Our careful analysis suggests that the **q$_1$**-modulation, with a wave vector of **q$_1$**= (1/3, 1/3, 2), can be interpreted by a 3-dimensional order of Fe$^{2+}$-Fe$^{2.5}$-Fe$^{3+}$. The **q$_2$** modulation in general is much weaker than the **q$_1$** modulation, and its average wave vectors at 20 K can be roughly written as **q$_2$**=(0 0 3/2)+ **q$_1$**/10. Further experimental measurements indicate that the **q$_2$** modulation also depends evidently on the local defective structures.

The origin of ferroelectricity in LuFe$_2$O$_4$ materials is considered to be essentially in correlation with the low-temperature CO. Fig. 6c shows a schematic charge ordered pattern in accordance with our experimental observations. It is remarkable that positive charges (Fe$^{3+}$ sites) and negative charges (Fe$^{2+}$ sites) are actually crystallized in parallel charge stripes going along the view direction. This charge stripe phase shows up a clear monoclinic feature with an evident electric polarity [20]. In order to facilitate the analysis on the electric polarization, we can also characterize this charge concentration in this stripe phase as a charge-density wave (CDW) in association with the $q_1$-modulation. This CDW running along the <116>-direction is not in a simple sinusoidal fashion, but being strongly affected by charge frustration as illustrated in Fig. 7a. It is easily visible that the average centers of $Fe^{2+}$ (negative) and $Fe^{3+}$ (positive) planes have a clear relative shift and directly results in a local electric polarization along the $q_1$ direction. It is worth to point out that the charge frustration could evidently affect the degeneracy of ground state in this system as similarly discussed for the spin frustration in triangular lattice [21], the presence of charge frustration of ground state of $LuFe_2O_4$ could yield a degenerate Fe site in an average valence of 2.5. On the other hand, if we analyzed this system based on $Fe^{2+}/Fe^{3+}$ ordering without considering the frustrated degeneracy ($Fe^{2.5+}$ state), a charge-density wave would behave in simple sinusoidal fashion, this kind of CO state cannot result in visible ferroelectricity due to coincidence of the centers of positive and negative charges in a supercell (i.e. A period CDW) as illustrated in Fig. 7b. We therefore conclude that the structural models simply base on $Fe^{2+}$ and $Fe^{3+}$ order contain essential difficulties to interpret the ferroelectricity in $LuFe_2O_4$ and also have an apparent contradiction with our low-temperature experimental results.

In the ferroelectric point of view, the frustrated type of charge configuration has the notable similarity with the coherent arrangement of electric dipoles discussed commonly in the conventional ferroelectric materials. Therefore, the $LuFe_2O_4$ crystal has a ferroelectric CO ground state in which the coherent arrangements of electric dipoles are realized by charge stripe order. It is noted that the parent phase of $LuFe_2O_4$ has the rhombohedral symmetry with the space group of R-3m, therefore, the CO modulation, together with the local electric polarization, could evenly appear in three crystallographically equivalent <116>-directions around the rhombohedra axis, and the resultant spontaneous electric polarization goes along the c-axis direction [20] in $LuFe_2O_4$, this conclusion is in good agreement with the major experimental data reported in previous literatures [2,22]. In order to directly observed the superstructures in three crystallographically equivalent <116>-directions, we have carried our systematical investigations by titling a crystal to the relevant orientations. Fig. 8 shows a series of diffraction patterns taken from a crystal with different orientations, illustrating the presence of superstructures along systematic {116} directions. We start from the [001] zone axis pattern, the [1-1-1] zone-axis diffraction pattern is reached after the specimen is tilted about 13° about 110 direction; the [241] zone-axis diffraction pattern is obtained after the specimen is tilted 25° along [2-10] direction. It is notable three diffraction patterns (position 3 in the Kikuchi map) show the similar superstructure spots corresponding with $q_1$ modulation.

In $LuFe_2O_4$ materials, two important phase transitions are proposed in high temperature range, i.e. the ferroelectric phase transition at about 350K and a charge order to disorder transition at about 530K. We therefore carried out an in-situ heating TEM observation to reveal the temperature dependence of the superstructure reflections in present system. Fig. 9a, b, c, and d shows a series diffraction patterns taken in the temperature range of 300K to 530K, illustrating the changes of superstructure streaks. The experimental results demonstrate that the intensities of the superstructure reflections decrease progressively with increasing temperature, and normally disappear above 530K. The disappearance of the diffuse lines directly suggests the electron charges in this material becomes disorder above this critical temperature. Fig. 9e shows the microphotometric density curves measured along a*+b* direction, clearly illustrating the decrease of the intensities of the superstructure peaks at high temperatures. TEM investigations also reveal that this transition is reversible, and the superstructure reflections becomes visible again as the temperature is lowered below 500K, careful measurements on the superstructure could reveal certain hysteretic behaviors, this fact suggests a complex nature for this phase transition in $LuFe_2O_4$ materials.

## Ⅳ CONCLUSIONS

In summary, the $LuFe_2O_4$ materials in general contain notable (001) twins along the c-axis direction, the $LuFe_2O_4$ crystals cross each boundary are rotated by 180° with respect to one another. The in-situ cooling TEM observations reveal remarkable temperature dependence of the superstructure modulations. Two structural modulations ($q_1$ and $q_2$) become clearly visible at the temperature of 20K. The $q_2$ modulation in general is much weaker than $q_1$. The $q_1$=(1/3 1/3 2) modulation can be well interpreted by the $Fe^{2+}$, $Fe^{3+}$, and $Fe^{2.5+}$ order in a stripe like configuration in the ground state. Moreover, it is noted that the ferroelectric phase transition and charge order-disorder transition occur in high temperature range in the $LuFe_2O_4$ materials; we therefore carried out an in-situ heating TEM observation to reveal the temperature dependence of the superstructure reflections. As a result, the intensities of superstructure reflections decrease gradually with the increase of temperature, and become invisible at around the 530K. Based on our systematic experimental results, we can conclude that the CO state in $LuFe_2O_4$ depends markedly on temperature between 530K and 20K, which plays a critical role for understanding the rich structural transitions and the electronic ferroelectricity observed in kind of materials.

1. N. Ikeda, *et al.*, *J. Phys. Soc. Jpn.* **64**, 1371 (1995).

2. N. Ikeda, *et al.*, *J. Phys. Soc. Jpn*. **69**, 1526 (2000).

3. N. Ikeda, *et al.*, *J. Phys. Soc. Jpn*. **63**, 4556 (1994).

4. K. Yoshiia, N. Ikeda, A. Nakamura, *Physica B* **378-380**, 585 (2006).

5. M.A. Subramanian, et al., *Adv. Mater.* **18**, 1737 (2006).

6. S-W. Cheong and M. Mostovoy, Nature Mater. **6**, 13(2007).

7. M. Fiebig, *et al.*, *Nature* **419**, 818 (2002).

8. B.B.V. Aken, T.T.M. Palstra, A. Filippetti, & N.A. Spaldin, *Nature Mater*. **3**, 164 (2004).

9. Y. Yamada, K. Kitsuda, S. Nohdo & N. Ikeda, *Phys. Rev. B* **62**, 12167 (2000).

10. J. Iide, *et al.*, *J. Phys. Soc. Jpn.* **62**, 1723 (1993).

11. N. Ikeda, *Physica B* **241-243**, 820 (1998).

12. Y. Yamada, & N. Ikeda, *J. Korean Phys. Soc.* **32**, S1 (1998).

13. R.J. Cava, A.P. Ramirez, Q. Huang, & J.J. Krajewski, *J. Solid. State Chem.* **140**, 337 (1998).

14. Y. Todatey, C. Kikutay, E. Himotoy, M. Tanakay & J. Suzukiz, J. Phys.: Condens. Matter **10,** 4057–4070 (1998).

15. K. Siratori, *et al.*, *J. Phys. Soc. Jpn.* **59**, 631 (1990).

16. H. Kito, *et al.*, *J. Phys. Soc. Jpn.* **64**, 2147 (1995).

17. M. Isobe, N. Kimizuka, J. Iida & S. Takekawa, Acta Cryst. C **46**, 1917 (1990).

18. Y. Horibe, K. Kishimoto, S. Mori & N. Ikeda, *J. Korean Phys. Soc.* **46**, 192 (2005).

19. Y. Yamada, S. Nohdo, & N. Ikeda, *J. Phys. Soc. Jpn.* **66**, 3733 (1997).

20. Y. Zhang, H. X. Yang, C. Ma, H. F. Tian & J. Q. Li (cond-mat/0702588, unpulished)

21. Mekata, M et al., J.phys. Soc.Jpn. **44**, 806(1978).

22. N. Ikeda, *et al.*, *Nature* **436**, 1136 (2005).

Figure captions

Fig. 1(a) X-ray diffraction (XRD) pattern from a LuFe$_2$O$_4$ sample. (b) A structural model schematically illustrating that Lu-O layer and Fe-O double layer stacking alternatively along the c axis, O-atoms are omitted for clarity. (c) A SEM image of LuFe$_2$O$_4$, clearly illustrating layered structural feature in this material.

Fig. 2 Electron diffraction patterns of LuFe$_2$O$_4$ taken along the (a) [1-1-1], (b) [33-1] and (c) [100], (d) [1-10] zone-axis directions at room temperature, respectively. All main diffraction spots can be well indexed by R-3m structure with lattice parameter of a=0.344nm, and c=2.5259nm. The superstructure streaks are clearly visible at the systematic position of (1/3 1/3 L).

Fig. 3 The [100] HRTEM image with a large magnification clearly displaying the atomic structure of LuFe$_2$O$_4$. Inset shows an image of theoretical simulations.

Fig. 4(a) Bright-field TEM image showing the twinning lamellae along c-axis direction. (b) , (c) and (d) show the [100] zone-axis SAD pattern of area Ⅰ Ⅱ and Ⅲ, respectively, clearly demonstrating twinning relationship across the boundary.

Fig. 5(a) HRTEM image showing the atomic structure nearby a boundary. The insets show the structure model for this twinning structure. (b) A HRTEM image with a large magnification clearly displaying a stacking fault on the boundary.

Fig. 6(a) [1-10] zone diffraction pattern at about 20K, demonstrating the presence of clear superstructure spots following the main diffraction spots. (b) A schematic illustration for the **q$_1$** and **q$_2$** modulation at low temperature. (c) Structural model schematically illustrating the ordered charge stripes of Fe$^{2+}$, Fe$^{3+}$ and Fe$^{2.5+}$ in the ground state.

Fig.7(a) The CDW in non-sinusoidal fashion corresponding with $q_1$-modulation. The resultant polarization is indicated. (b) A CDW in a sinusoidal fashion, no polarization is expected due to the coincidence of the positive and negative charge centers for the supercell.

Fig.8 A series of low temperature diffraction patterns illustrating the structural features along several major zone axes. It is notable that three diffraction patterns (position 3 in the Kikuchi map) show the similar superstructure spots along three crystallographically equivalent <116>-directions.

Fig. 9 The temperature dependence of the superstructure reflections from 300K to 530K. The intensities of superlattice reflections become weaker and more diffuse when the temperature is increased, and disappear above 530K.

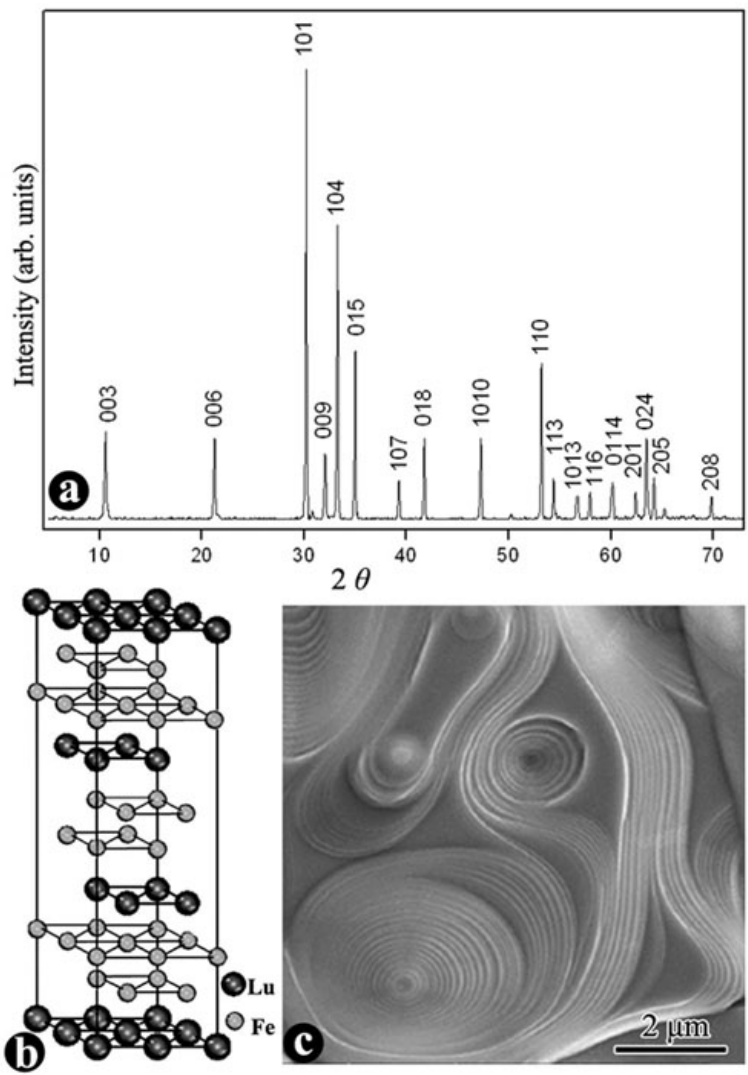

**Fig. 1**

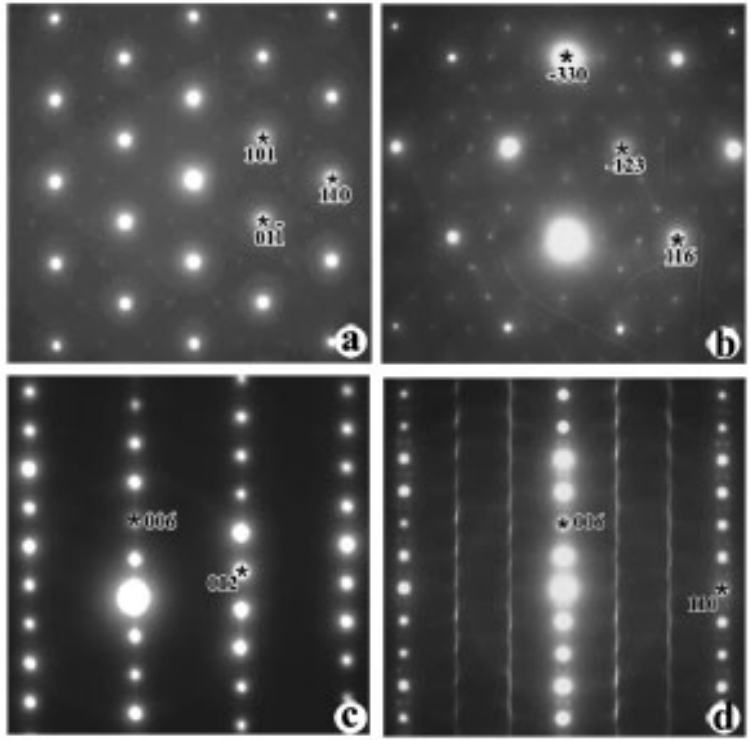

**Fig. 2**

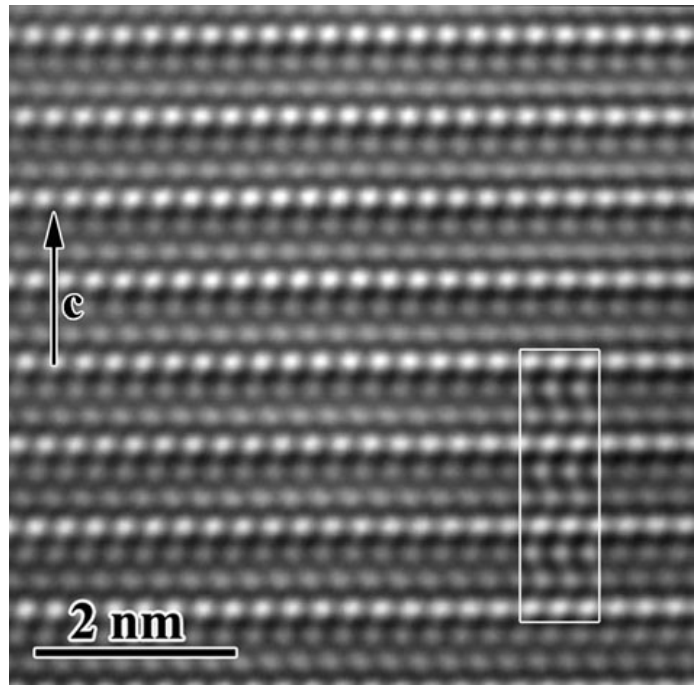

**Fig. 3**

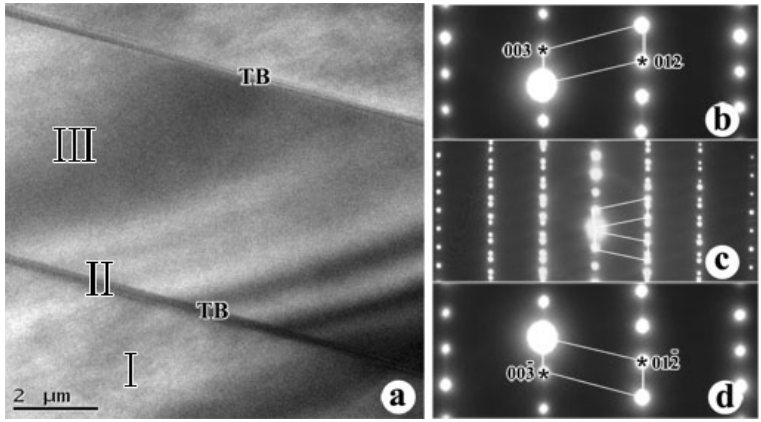

**Fig. 4**

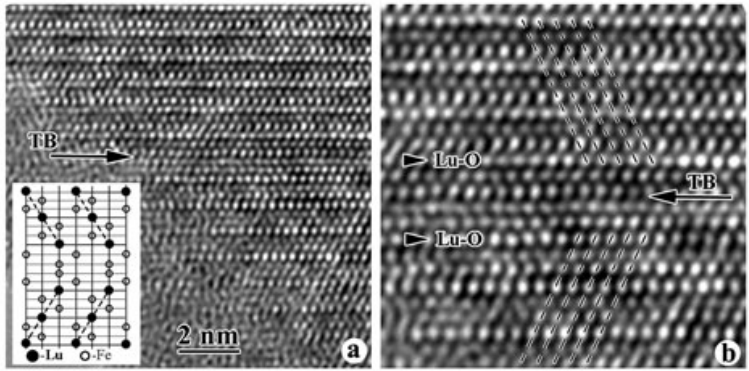

**Fig. 5**

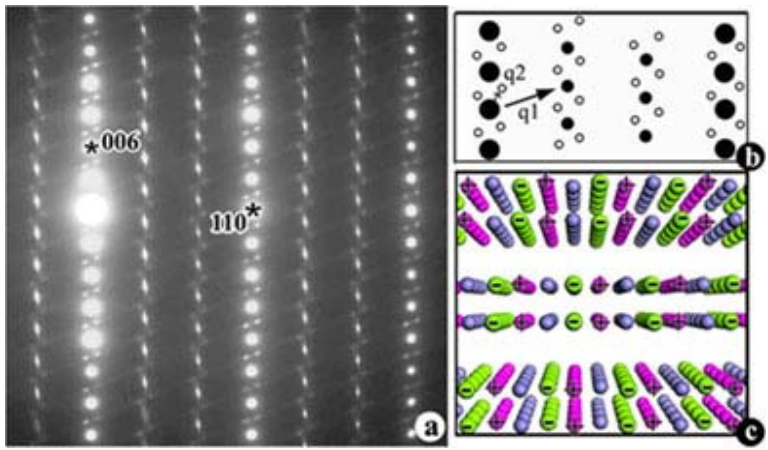

**Fig. 6**

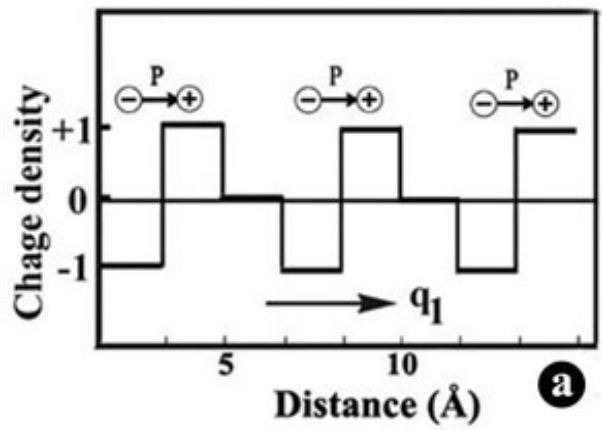

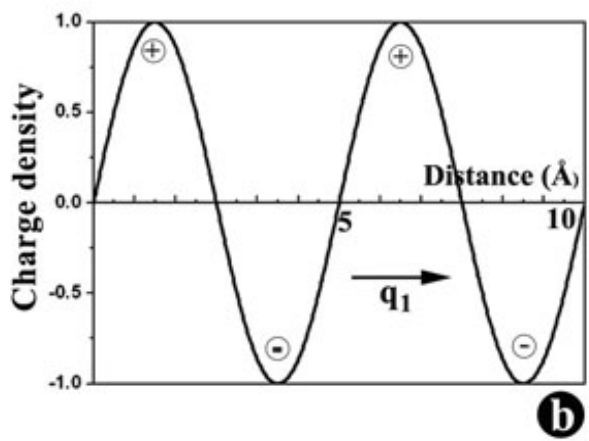

**Fig. 7**

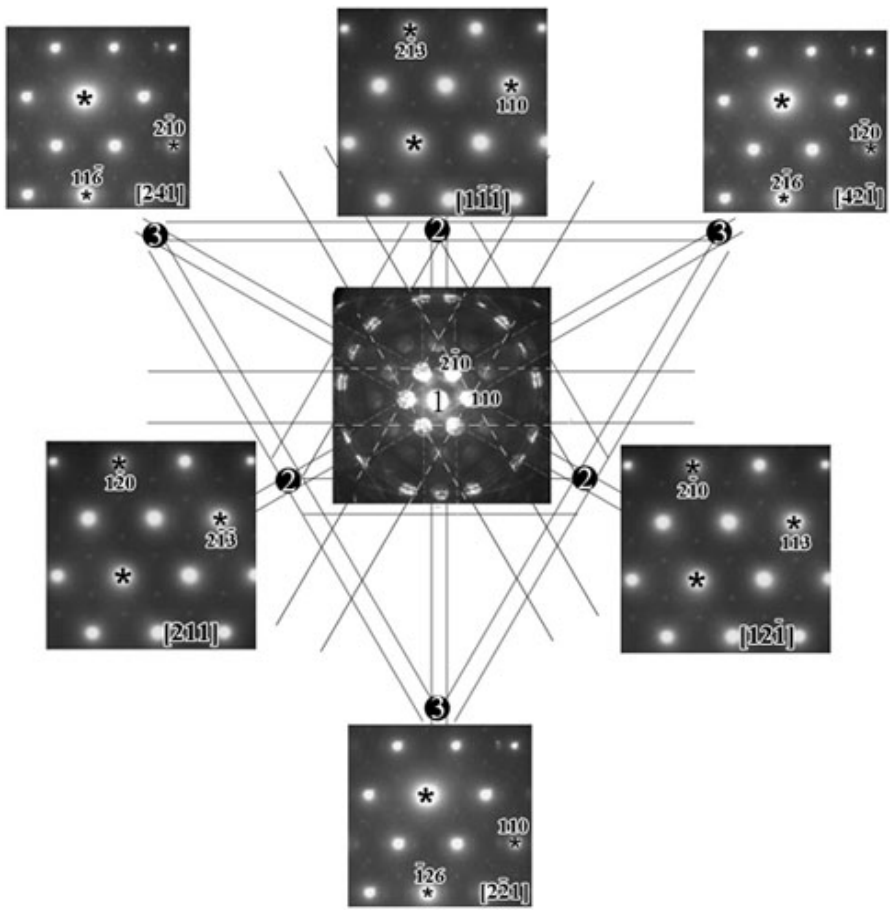

**Fig. 8**

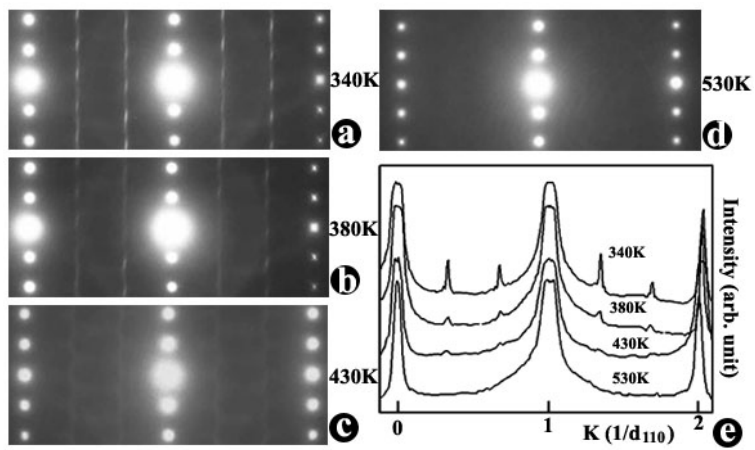

Fig. 9